\documentclass[11pt]{article}
\usepackage{mathpazo}
\usepackage[T1]{fontenc}
\usepackage[latin9]{inputenc}
\usepackage{geometry}
\usepackage{xcolor}
\usepackage{pdfcolmk}
\usepackage{array}
\usepackage{booktabs}
\usepackage{textcomp}
\usepackage{amsmath}
\usepackage{amssymb}
\usepackage{graphicx}
\PassOptionsToPackage{normalem}{ulem}
\usepackage{ulem}
\usepackage[unicode=true,
 bookmarks=true,bookmarksnumbered=true,bookmarksopen=false,
 breaklinks=true,pdfborder={0 0 1},backref=false,colorlinks=true]
 {hyperref}
\hypersetup{
 urlcolor=urlcolor,linkcolor=linkcolor,citecolor=citecolor}

\makeatletter

\providecommand{\tabularnewline}{\\}
\providecolor{lyxadded}{rgb}{0,0,1}
\providecolor{lyxdeleted}{rgb}{1,0,0}
\DeclareRobustCommand{\lyxsout}[1]{\ifx\\#1\else\sout{#1}\fi}

\definecolor{linkcolor}{rgb}{0,0,1}
\definecolor{citecolor}{rgb}{0,0,1}

\def\lt{<}

\usepackage[nomarkers,tablesfirst]{endfloat}

\usepackage{bookmark}
\usepackage{cancel}
\usepackage[english]{babel}
\makeatother

\begin{document}
\title{{\Large{}Method of Exact Solutions Code Verification of a Superelastic
Constitutive Model in a Commercial Finite Element Solver}}
\author{Kenneth I. Aycock\thanks{Division of Applied Mechanics, Office of Science and Engineering Laboratories,
Center for Devices and Radiological Health, United States Food and
Drug Administration, 10903 New Hampshire Avenue, Silver Spring, MD
20993}\and Nuno Rebelo\thanks{Nuno Rebelo Associates, LLC, Fremont, CA 94539}\and Brent
A. Craven\textsuperscript{{*}}}

\maketitle
\vspace{-25pt}

\begin{abstract}
The superelastic constitutive model implemented in the commercial
finite element code \texttt{ABAQUS} is verified using the method of
exact solutions (MES). An analytical solution for uniaxial strain
is first developed under a set of simplifying assumptions including
von Mises-like transformation surfaces, symmetric transformation behavior,
and monotonic loading. Numerical simulations are then performed, and
simulation predictions are compared to the exact analytical solutions.
Results reveal the superelasticity model agrees with the analytical
solution to within one ten-thousandth of a percent (0.0001\%) or less
for stress and strain quantities of interest when using displacement-driven
boundary conditions. Full derivation of the analytical solution is
provided in an Appendix, and simulation input files and post-processing
scripts are provided as supplemental material.
\end{abstract}

\section{Introduction}

Superelastic nickel titanium (nitinol) alloys are commonly used in
medical devices such as guidewires, dental arches, and self-expanding
peripheral stents, stent grafts, heart valve frames, and inferior
vena cava filters. Because of nitinol's unique material behavior and
the complex geometry of most nitinol devices, engineers and scientists
often use physics-based computational modeling and simulation to predict
device mechanics and fatigue safety factors as part of non-clinical
bench performance testing. As described in ASME V\&V40-2018 \cite{v2018assessing},
model predictions relied on for decision making should be accompanied
by verification and validation (V\&V) evidence demonstrating simulation
credibility commensurate with the risk associated with the intended
model use. However, rigorous code verification evidence for medical
device simulations is often omitted (e.g., see Figure 1 in \cite{Baumann_2021}),
in part due to the lack of detailed examples to facilitate these studies.
Herein, we aim to provide such an example for superelastic nitinol.

In previous work, we demonstrated gold-standard method of manufactured
solutions (MMS) code verification of the commercial finite element
software \texttt{ABAQUS} for various linear and nonlinear elastostatics
problems \cite{aycock2020method}. However, direct MMS verification
of the superelastic model commonly used to simulate nitinol was not
possible due to the lack of a closed-form representation of the underlying
rate- and history-dependent constitutive equations. An approach recommended
in the literature for rigorously verifying similarly complex, plasticity-based
constitutive models is to perform method of exact solutions (MES)
verification on an affine deformation problem with prescribed strains
or displacements \cite{kamojjala2015verification}. In this study,
we perform MES code verification of the superelastic constitutive
model in \texttt{ABAQUS}.

\section{Methods}

\subsection{Constitutive model summary}

\begin{figure}
\centering{}\includegraphics[width=6.5in]{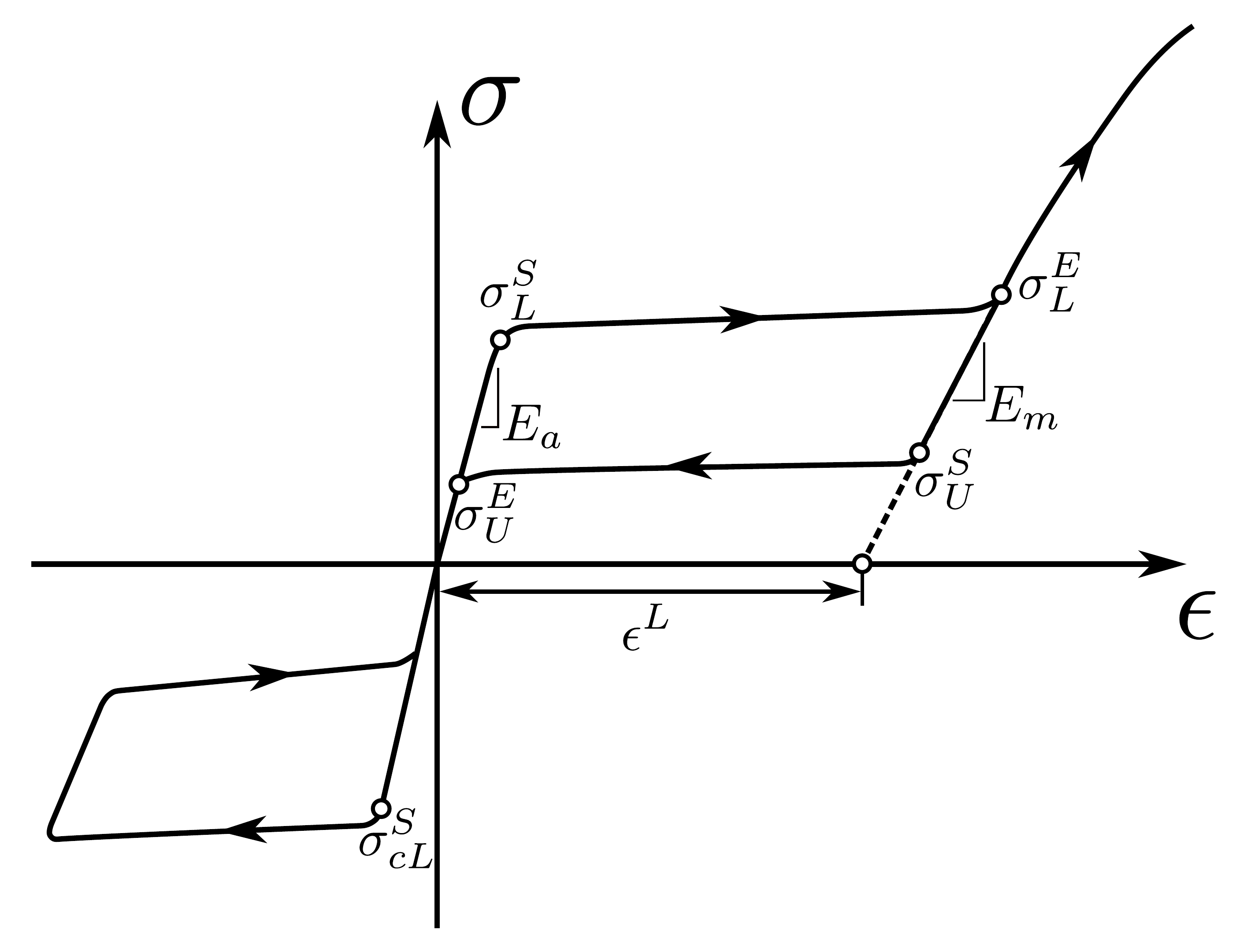}\caption{Notional uniaxial stress-strain curve for superelasticity \texttt{UMAT}
in \texttt{ABAQUS/Standard}.\label{fig:NiTi-curve}}
\end{figure}

Superelastic constitutive behavior was first implemented in \texttt{ABAQUS/Standard}
as a user-material (\texttt{UMAT}) by Rebelo et al \cite{rebelo2000finite,rebelo2001simulation}
in 2000. In brief, the model is based on the work of Auricchio and
Taylor \cite{auricchio1997shape,auricchioTaylorLubliner1997} and
leverages generalized plasticity theory to model the dependency of
the material stiffness on the current stress state. More specifically,
the model uses a mixture-based approach to simulate the stress-induced
solid-solid phase transformation between cubic (B2) austenite and
monoclinic (B19\textquoteright ) martensite, tracked by the martensite
fraction parameter $\zeta$. Additional details on the constitutive
model and the associated transformation flow rule are provided in
the \nameref{sec:Appendix}. A notional stress-strain response and
associated input parameters are summarized in Figure~\ref{fig:NiTi-curve}
and Table~\ref{tab:UMAT-parameters}, respectively.
\begin{table}
\caption{Input parameters for superelastic \texttt{UMAT} in \texttt{ABAQUS/Standard}.\label{tab:UMAT-parameters}}

\centering{}%
\begin{tabular}{c>{\centering}p{2.5in}}
\toprule 
parameter & description\tabularnewline
\midrule 
$E_{a}$ & Young's modulus, austenite\tabularnewline
$\nu_{a}$ & Poisson's ratio, austenite\tabularnewline
$E_{m}$ & Young's modulus, martensite\tabularnewline
$\nu_{m}$ & Poisson's ratio, martensite\tabularnewline
$\epsilon^{L}$ & transformation strain\tabularnewline
$\left(\frac{\partial\sigma}{\partial T}\right)_{L}$ & change in transformation stresses with temperature, loading\tabularnewline
$\sigma_{L}^{S}$ & start of transformation, loading\tabularnewline
$\sigma_{L}^{E}$ & end of transformation, loading\tabularnewline
$T_{0}$ & reference temperature\tabularnewline
$\left(\frac{\partial\sigma}{\partial T}\right)_{U}$ & change in transformation stresses with temperature, unloading\tabularnewline
$\sigma_{U}^{S}$ & start of transformation, unloading\tabularnewline
$\sigma_{U}^{E}$ & end of transformation, unloading\tabularnewline
$\sigma_{cL}^{S}$ & start of transformation, compression\tabularnewline
$\epsilon_{V}^{L}$ & volumetric transformation strain\tabularnewline
\bottomrule
\end{tabular}
\end{table}

\subsection{Simplifying assumptions\label{subsec:Simplifying-assumptions}}

\begin{figure}
\begin{centering}
\includegraphics[width=6.5in]{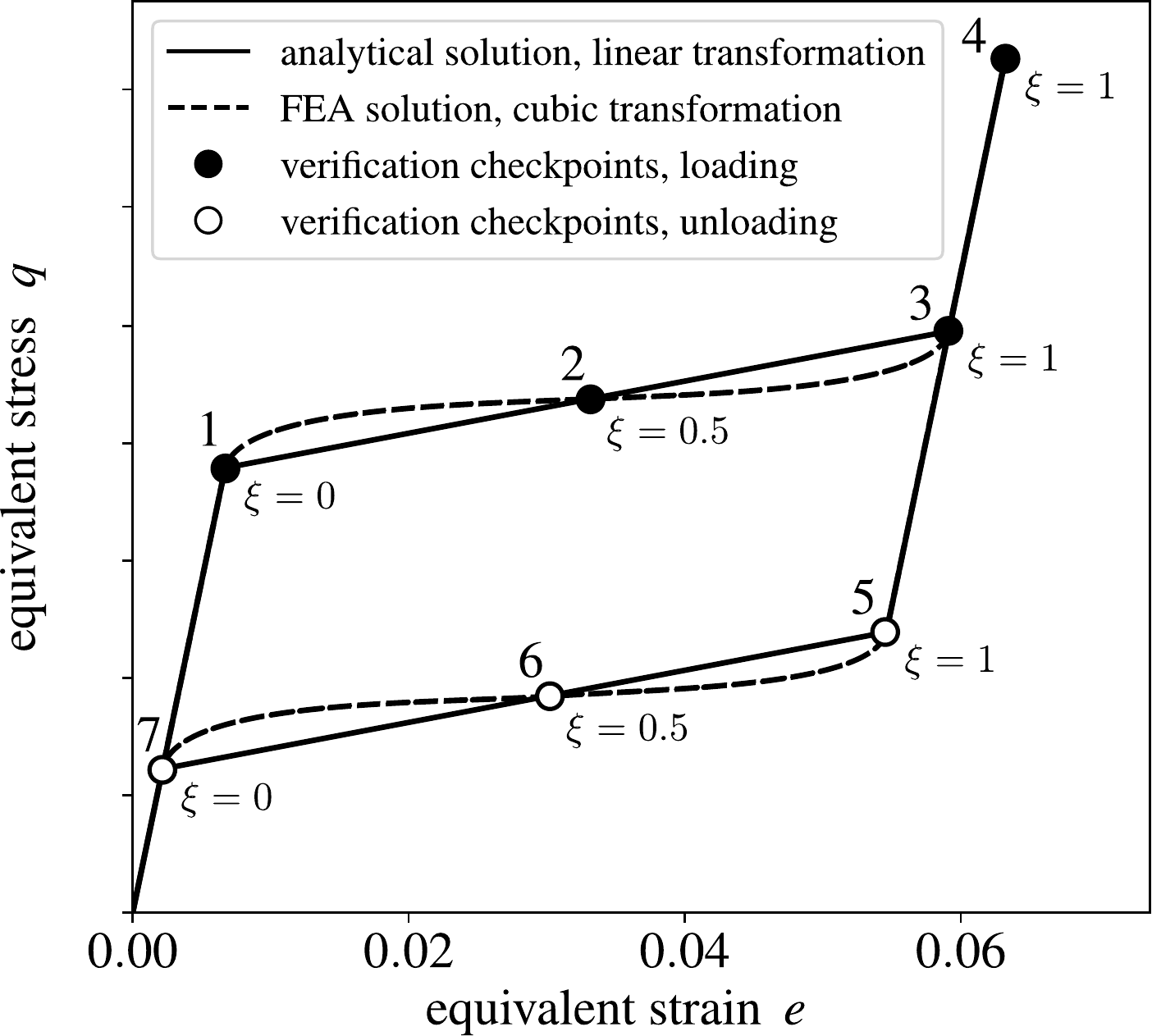}
\par\end{centering}
\caption{Illustration comparing the linear and nonlinear transformation solutions.
Markers indicate locations at the beginning (1,5), mid-point (2,6),
and end (3,7) of the loading and unloading transformation plateaus
where the solutions are equivalent, and $\xi$ indicates the martensite
fraction. The cubic shape of the nonlinear curve is exaggerated to
facilitate visualization.\label{fig:NiTi-linear-vs-cubic-transformation}}
\end{figure}

The superelastic constitutive model in \texttt{ABAQUS} uses pressure-dependent
Drucker--Prager-like transformation surfaces and cubic transformation
equations to define nonlinear hardening during phase transformation
between austenite and martensite. As such, a general analytical solution
to the associated rate equations is not easily obtained and to our
knowledge has not been derived. Here, we instead derive an analytical
solution for linear transformation behavior. Because of the way the
nonlinear transformation equations are defined in \texttt{ABAQUS},
the linear and nonlinear transformation solutions should be equivalent
at the beginning, mid-point, and end of both the loading (austenite~$\rightarrow$~martensite)
and unloading (martensite~$\rightarrow$~austenite) transformations
under the following assumptions (see Figure~\ref{fig:NiTi-linear-vs-cubic-transformation}
and Table~\ref{tab:verification-points}):
\begin{itemize}
\item symmetric transformation behavior in tension and compression, i.e.
$\sigma_{L}^{S}=\sigma_{cL}^{S}$ (constitutive model becomes von
Mises-like rather than Drucker--Prager-like)
\item constant temperature (isothermal)
\item equal elastic moduli for austenite and martensite, i.e. $E_{a}=E_{m}$
\item pseudo-plasticity/superelasticity behavior only (i.e., the model is
not superelastic-plastic)
\item monotonic and proportional (i.e., radial) loading.
\end{itemize}
\begin{table}
\caption{Pairs of Mises equivalent stresses $q$ and martensite fractions $\xi$
defining verification points where the linear and nonlinear transformation
solutions are equivalent. Note each combination of $q$ and $\xi$
defines a unique location in Figure \ref{fig:NiTi-linear-vs-cubic-transformation}.\label{tab:verification-points}}

\centering{}%
\begin{tabular}{ccccl}
\toprule 
 &  & description & $q$ & $\xi$\tabularnewline
\midrule 
loading & 1 & beginning of transformation & $\sigma_{L}^{S}$ & 0.0\tabularnewline
 & 2 & mid-point of transformation & $\frac{\sigma_{L}^{S}+\sigma_{L}^{E}}{2}$ & 0.5\tabularnewline
 & 3 & end of transformation & $\sigma_{L}^{E}$ & 1.0\tabularnewline
 & 4 & end of loading (pure martensite) & $q_{\textrm{max}}$ & 1.0\tabularnewline
unloading & 5 & beginning of transformation & $\sigma_{U}^{S}$ & 1.0\tabularnewline
 & 6 & mid-point of transformation & $\frac{\sigma_{U}^{S}+\sigma_{U}^{E}}{2}$ & 0.5\tabularnewline
 & 7 & end of transformation & $\sigma_{U}^{E}$ & 0.0\tabularnewline
\bottomrule
\end{tabular}
\end{table}
With these assumptions, we derive an exact analytical solution for
the uniaxial strain of a single cubic element undergoing linear transformation
and monotonic loading.

\subsection{Problem description\label{subsec:Problem-setup}}

\begin{figure}
\centering{}\includegraphics[width=4.5in]{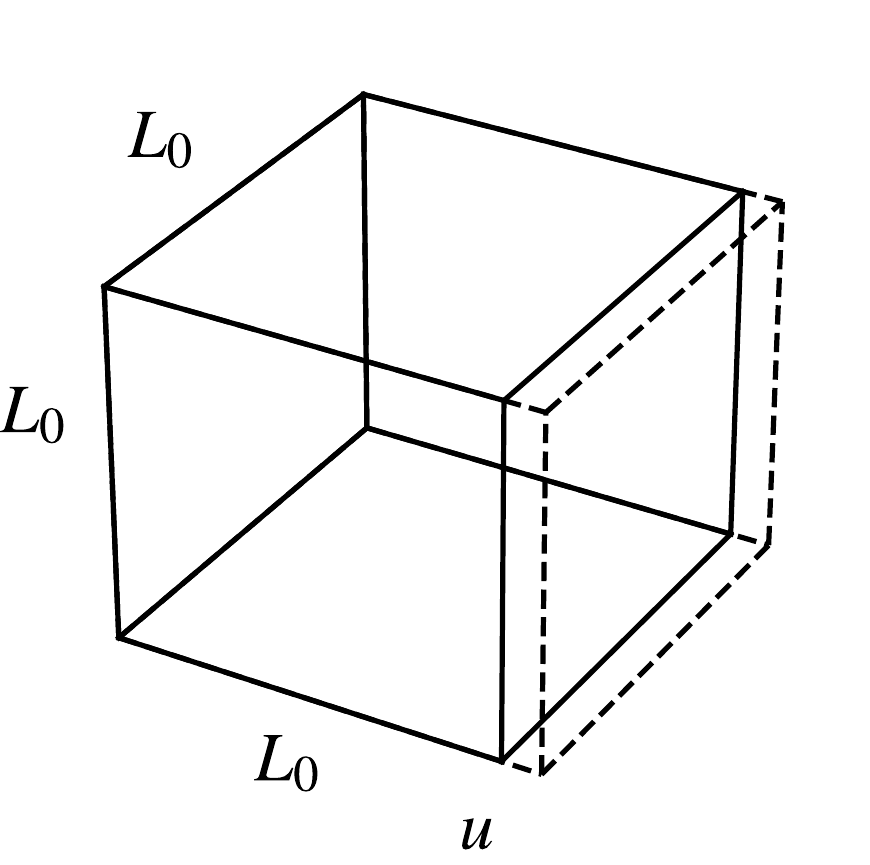}\caption{Schematic showing uniaxial strain of a unit cube with starting side
length $L_{0}$ and final length $L_{F}=L_{0}+u$.\label{fig:unit-cube}}
\end{figure}

Uniaxial strain of a unit cube is considered, i.e.,
\begin{align}
\epsilon_{11} & =f(t)\label{eq:uniaxial-strain-e11}\\
\epsilon_{22} & =\epsilon_{33}=\epsilon_{12}=\epsilon_{13}=\epsilon_{23}=0\label{eq:uniaxial-strain}
\end{align}
where $\epsilon_{ij}$ are the components of the logarithmic strain
tensor and $t$ is the simulation pseudo-time for the elastostatic
analysis. The cube has an initial side length $L_{0}=1$ and final
side length in the direction of the applied strain of $L_{F}=L_{0}+u$,
where $u$ is the applied displacement (Figure~\ref{fig:unit-cube}).
An unloading analysis is additionally performed to return the single
element to the reference configuration.

During the loading and unloading analyses, the current length $l$
is defined as a linear ramping between $L_{0}$ and $L_{F}$,
\begin{equation}
l=\begin{cases}
L_{0}+u\,t & \textrm{loading}\\
L_{0}+u\left(1-t\right) & \textrm{unloading}
\end{cases}\label{eqn:l}
\end{equation}
with pseudo-time $t$ in the range 
\begin{equation}
0<t<1\,.
\end{equation}
We can also quantify the deformation using the non-dimensional stretch,

\begin{equation}
\lambda=\frac{l}{L_{0}}=\begin{cases}
1+\frac{u\,t}{L_{0}} & \textrm{loading}\\
1+\frac{u\left(1-t\right)}{L_{0}} & \textrm{unloading.}
\end{cases}\label{eqn:stretch}
\end{equation}

\subsection{Analytical solution}

The exact analytical solution is derived in detail in the \nameref{sec:Appendix}.
In summary, given
\begin{itemize}
\item Mises stress $q$ and martensite fraction $\xi$ at the verification
points in Table~\ref{tab:verification-points},
\item shear and bulk moduli $G$ and $K$,
\item assumptions from Section \ref{subsec:Simplifying-assumptions},
\end{itemize}
the analytical solutions for pseudo-time and the remaining field variables
are summarized in Table~\ref{tab:analytical-solution-summary}.
\begin{table}
\centering{}\caption{Analytical solutions for key quantities of interest at verification
points $\left(q,\,\xi\right)$ listed in Table~\ref{tab:verification-points}
and illustrated in Figure \ref{fig:NiTi-linear-vs-cubic-transformation}.\label{tab:analytical-solution-summary}}
\begin{tabular}{ccc}
\toprule 
description & parameter & expression\tabularnewline
\midrule
total logarithmic strain & $\epsilon$ & $\frac{q}{2G}+\frac{3}{2}\epsilon^{L}\xi$\tabularnewline
hydrostatic pressure & $p$ & $-K\epsilon$\tabularnewline
deviatoric stress tensor & $\mathbf{S}$ & $\frac{2}{3}q\,\mathbf{n}$\tabularnewline
Cauchy stress tensor & $\boldsymbol{\sigma}$ & $\mathbf{S}-p\mathbf{I}$\tabularnewline
stretch & $\lambda$ & $\exp\left(\epsilon\right)$\tabularnewline
simulation pseudo-time & $t$ & $\begin{cases}
\frac{L_{0}(\lambda-1)}{u} & \text{loading}\\
1-\frac{L_{0}(\lambda-1)}{u} & \text{unloading}
\end{cases}$\tabularnewline
\bottomrule
\end{tabular}
\end{table}

\subsection{Numerical simulations}

Single-element verification simulations were performed in \texttt{ABAQUS/Standard}
versions R2016x and 2022 using a single core of an Intel Xeon E5-4627
v4 processor. Commonly used continuum element types \texttt{C3D8,
C3D8R, C3D8I}, \texttt{C3D20}, and \texttt{C3D20R} are investigated
in both software versions. Material constants are prescribed as defined
in Table \ref{tab:Input-variables}. The input files for all cases
are provided in supplemental material.

\texttt{ABAQUS} solves nonlinear mechanics problems in an incremental
fashion for each simulation ``step'' or load sequence. For the present
verification exercise, multiple steps are defined to facilitate extraction
of field outputs at the precise pseudo-times where the analytical
and numerical solutions agree if the constitutive code is correctly
implemented (Tables~\ref{tab:verification-points}~and~\ref{tab:analytical-solution-summary}).
Displacement-controlled boundary conditions are prescribed to enforce
the uniaxial strain conditions described in Eqn. \ref{eq:uniaxial-strain-e11}--\ref{eqn:l}.
Increment sizes are defined such that each step consists of 100 increments.
Numerical calculations of Mises stress $q$, uniaxial strain $\epsilon_{11}$,
and stress components $\sigma_{11}$, $\sigma_{22}$, and $\sigma_{33}$
are extracted at each pseudo-time point using an \texttt{ABAQUS/Python}
script. Note these results could also be obtained from \texttt{ABAQUS}
printed output provided such output is requested at appropriate simulation
pseudo-times. For elements with multiple integration points, quantities
of interest are averaged across all integration points. Percentage
error magnitudes between the numerical and analytical solutions are
calculated as $|\phi_{\textrm{numerical}}-\phi_{\textrm{analytical}}|/\phi_{\textrm{analytical}}\times100$
for each quantity of interest $\phi$.

A second set of single-element verification simulations using traction
boundary conditions was also performed. In brief, the moving displacement
boundary condition used above was replaced with a pressure boundary
condition, with the prescribed pressure equal to the negative of the
$\sigma_{11}$ component of the Cauchy stress tensor from the analytical
solution (Table \ref{tab:analytical-solution-summary}).
\begin{table}
\caption{Input variables for single-element verification problem.\label{tab:Input-variables}}

\centering{}%
\begin{tabular}{cccc}
\toprule 
description & parameter & value & units\tabularnewline
\midrule 
initial length & $L_{0}$ & 1 & mm\tabularnewline
maximum displacement & $u$ & 0.1 & mm\tabularnewline
shear modulus & $G$ & 19,000 & MPa\tabularnewline
bulk modulus & $K$ & 42,000 & MPa\tabularnewline
\midrule 
Young's modulus & $E_{\left\{ a,m\right\} }$ & $\frac{9KG}{3K+G}$ & MPa\tabularnewline
Poisson's ratio & $\nu_{\left\{ a,m\right\} }$ & $\frac{3K-2G}{2(3K+G)}$ & --\tabularnewline
transformation strain & $\epsilon^{L}=\epsilon_{V}^{L}$\textsuperscript{{*}} & 0.05 & --\tabularnewline
change in transformation stress with temperature & $\left(\frac{\partial\sigma}{\partial T}\right)_{\{L,U\}}$ & 0 & $\frac{\textrm{MPa}}{{^\circ}C}$\tabularnewline
start of transformation, loading & $\sigma_{L}^{S}=\sigma_{cL}^{S}$ & 370 & MPa\tabularnewline
end of transformation, loading & $\sigma_{L}^{E}$ & 410 & MPa\tabularnewline
reference temperature & $T_{0}$ & 0 & °C\tabularnewline
start of transformation, unloading & $\sigma_{U}^{S}$ & 160 & MPa\tabularnewline
end of transformation, unloading & $\sigma_{U}^{E}$ & 120 & MPa\tabularnewline
\midrule
\multicolumn{4}{l}{{\footnotesize{}}\textsuperscript{{\footnotesize{}{*}}}{\footnotesize{}The
input of a desired volumetric transformation strain $\epsilon_{V}^{L}$
in ABAQUS other than $\epsilon^{L}$ toggles}}\tabularnewline
\multicolumn{4}{l}{{\footnotesize{}the use of a non-associated flow rule. Because the
model herein simplifies to Mises-equivalent}}\tabularnewline
\multicolumn{4}{l}{{\footnotesize{}transformation, the volumetric transformation strain
is automatically zero.}}\tabularnewline
\end{tabular}
\end{table}

\section{Results}

\begin{figure}
\centering{}\includegraphics[width=6.5in]{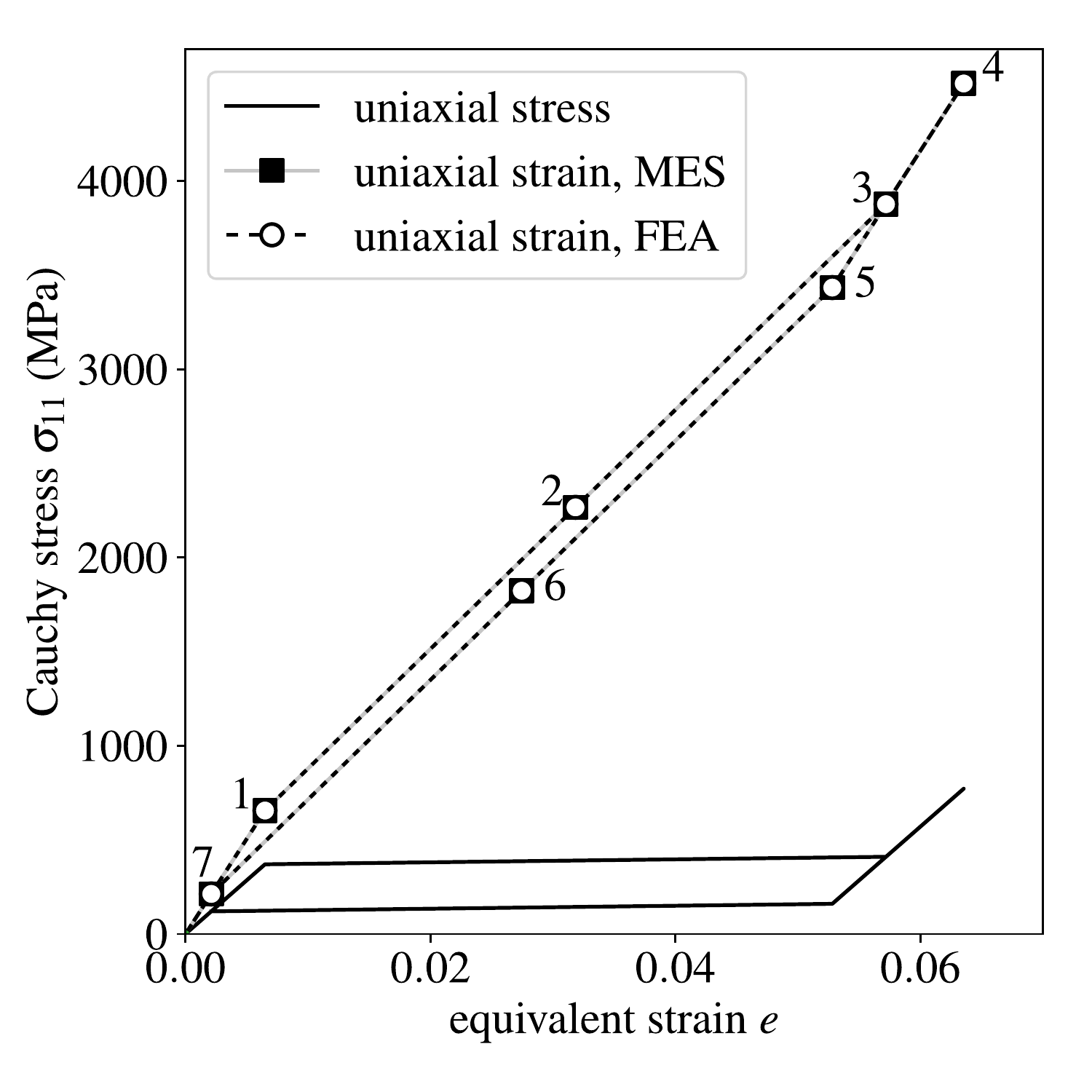}\caption{Comparison of the numerical and analytical results under uniaxial
strain conditions. The corresponding superelastic stress-strain curve
under uniaxial stress is also shown for reference.\label{fig:MES-results}}
\end{figure}

Each single-element simulation was completed within approximately
30 seconds with no clear influence of element type or solver version
on simulation time (25.7~$\pm$~1.3~sec and 35.6~$\pm$~3.4~sec
for displacement- and traction-controlled simulations, respectively).
Simulations converged at each pseudo-time increment without the need
for decreasing the increment size. Qualitatively, the numerical and
analytical solutions are identical across all element types and software
versions considered (Figure~\ref{fig:MES-results}). The loading
and unloading plateaus exhibit relatively large stiffnesses under
uniaxial strain conditions compared to those associated with uniaxial
stress conditions (Figure~\ref{fig:MES-results}). Indeed, the $\sigma_{11}$
Cauchy stresses reach values nearly an order of magnitude larger under
uniaxial strain conditions.

Quantitatively, for the displacement-controlled simulations, the results
are relatively insensitive to the continuum element type or software
version used, and extracted stress and strain measures are equivalent
to within six to eight significant figures (see Supplemental Materials).
One exception is the final equilibrium state after unloading where
small residual stresses and strains are observed (Table~\ref{tab:numerical-stress-strain}).
Comparing the analytical and numerical results, maximum errors in
stress and strain are on the order of one ten-thousandth of a percent
(Table~\ref{tab:numerical-stress-strain}). Although differences
in the percent error magnitudes are observed among the various solver
versions and element types investigated, the differences are relatively
small. Accordingly, only the largest error magnitudes are reported
in Table \ref{tab:numerical-stress-strain} for brevity.

Results are similar for traction-controlled conditions, although percent
error magnitudes increase to approximately 1e-03 at the end of transformation
in both loading and unloading. One exception is encountered using
\texttt{C3D8I }elements, where percent error magnitudes reach approximately
1\% at the end of transformation in unloading, and finite residual
stresses and strains are observed at the end of unloading. For full
details, see the supplemental material.
\begin{table}
\centering{}\caption{Numerical quantities of interest and percent error magnitudes calculated
at the verification time points using displacement-controlled boundary
conditions. The final row shows the largest deviations from the anticipated
zero-stress and zero-strain condition across the simulations performed.
The final three columns likewise report the maximum percent error
magnitudes (\% err.) observed across all simulations comparing the
numerical and analytical solutions. Percent error magnitudes are omitted
for the last row given division by zero. Stresses are in units of
megapascals (MPa). For full results using each solver version and
element type, see supplemental material.\label{tab:numerical-stress-strain}\smallskip{}
}
\begin{tabular}{ccccccccc}
\toprule 
{\small{}step} & {\small{}$t$} & {\small{}$q$} & {\small{}$\epsilon_{11}$} & {\small{}$\sigma_{11}$} & {\small{}$\sigma_{\{22,33\}}$} & {\small{}$\epsilon_{11}$ \% err.} & {\small{}$\sigma_{11}$ \% err.} & {\small{}$\sigma_{\{22,33\}}$ \% err.}\tabularnewline
{\small{}1} & {\small{}0.09784} & {\small{}370} & {\small{}0.009737} & {\small{}655.6} & {\small{}285.6} & {\small{}3.806e-06} & {\small{}3.267e-06} & {\small{}3.187e-06}\tabularnewline
{\small{}2} & {\small{}0.4892} & {\small{}390} & {\small{}0.04776} & {\small{}2266} & {\small{}1876} & {\small{}5.419e-07} & {\small{}6.237e-06} & {\small{}1.027e-06}\tabularnewline
{\small{}3} & {\small{}0.8958} & {\small{}410} & {\small{}0.08579} & {\small{}3876} & {\small{}3466} & {\small{}2.121e-06} & {\small{}4.420e-07} & {\small{}4.942e-07}\tabularnewline
{\small{}4} & {\small{}1} & {\small{}771.8} & {\small{}0.09531} & {\small{}4518} & {\small{}3746} & {\small{}6.165e-06} & {\small{}7.726e-06} & {\small{}3.672e-06}\tabularnewline
{\small{}5} & {\small{}0.1757} & {\small{}160} & {\small{}0.07921} & {\small{}3434} & {\small{}3274} & {\small{}1.168e-06} & {\small{}4.990e-07} & {\small{}5.234e-07}\tabularnewline
{\small{}6} & {\small{}0.5796} & {\small{}140} & {\small{}0.04118} & {\small{}1823} & {\small{}1683} & {\small{}3.009e-06} & {\small{}5.874e-06} & {\small{}6.362e-06}\tabularnewline
{\small{}7} & {\small{}0.9684} & {\small{}120} & {\small{}0.003158} & {\small{}212.6} & {\small{}92.63} & {\small{}1.078e-04} & {\small{}1.088e-04} & {\small{}1.179e-04}\tabularnewline
{\small{}8} & {\small{}1} & {\small{}3.796e-04} & {\small{}1.000e-08} & {\small{}6.739e-04} & {\small{}2.943e-04} & {\small{}N/A} & {\small{}N/A} & {\small{}N/A}\tabularnewline
\bottomrule
\end{tabular}
\end{table}

\section{Discussion and Conclusions}

Using the method of exact solutions, we demonstrate verification of
the superelastic constitutive model implemented in the \texttt{ABAQUS/Standard}
implicit finite element solver. Specifically, uniaxial strain conditions
are used to facilitate derivation of a closed-form solution for monotonic
loading and unloading through the full range of transformation behavior.
Although the uniaxial conditions are relatively simple to implement,
they generate a nontrivial stress state that differs from the uniaxial
stress condition typically used for model calibration. The verification
exercise is performed by extracting quantities of interest at specific
simulation increments where the numerical and analytical solutions
should theoretically agree if the superelastic model is properly implemented.
Simulation results reveal maximum errors in quantities of interest
are on the order of one ten-thousandth of a percent, providing evidence
that the model is indeed correctly implemented. The results are quantitatively
similar for all solver versions and continuum element types investigated.

Note that, while we observe relatively small error magnitudes, the
errors exceed machine precision for the double-precision floating-point
operations performed by \texttt{ABAQUS/Standard}. A potential explanation
is numerical incrementation. Each \texttt{ABAQUS} solution is computed
in 800 increments, and there is a potential for errors to be generated
at each increment. The results do not indicate obvious trends in error
magnitude with increasing pseudo-time increment during loading. Accordingly,
errors generated through incrementation are either negligible compared
to the overall observed error magnitudes, or they are offset by subsequent
increments such that they do not accumulate in simulation pseudo-time.
In contrast, error magnitudes do increase i)~during unloading when
using \texttt{C3D8I} elements in the 2022 solver under displacement
control, and ii)~at the end of transformation for all element types
and solver versions under traction control (see supplemental material),
possibly due to incrementation error. The largest error magnitudes
are observed at the end of unloading when using the \texttt{C3D8I}
elements under traction control. Based on further investigation, this
latter observation is uniquely associated with a combination of \texttt{C3D8I}
elements, traction boundary conditions, and the superelastic model
and is believed to be caused by the particular formulation of the
incompatible mode element type.

A few limitations should be noted. First, although the fully integrated
and simplified closed-form solution provided in Table \ref{tab:analytical-solution-summary}
is convenient for performing the verification exercise, the solution
is limited to strictly monotonic loading. Alternatively, the solution
could be generalized to consider the activation and evolution of the
martensite fraction $\xi$ under other loading paths, for example,
unloading from the midpoint on the upper plateau (i.e., between verification
points 2 and 6 in Figure \ref{fig:NiTi-linear-vs-cubic-transformation}).
Second, a number of simplifying assumptions were used to derive the
closed-form analytical solution. Consideration of tension-compression
asymmetry, differences in austenite and martensite elastic moduli,
and extensions of the superelasticity model such as the superelastic-plastic
implementation could be investigated in future work. Third, the verification
tests considered herein only address three-dimensional continuum elements
using the implicit solver in \texttt{ABAQUS}. Verification using other
element (e.g., beam and tetrahedral) and solver types (e.g., \texttt{ABAQUS/Explicit})
could be investigated in future work.

In conclusion, method of exact solutions code verification of the
superelasticity model in\texttt{ ABAQUS} was successful under uniaxial
strain conditions. Code verification evidence like that generated
by this study, alongside solution verification, experimental validation,
and uncertainty quantification evidence, are useful to support the
credibility of computational models, especially when model predictions
are used to inform high-risk decision making. To facilitate reproducibility
of this study using other hardware systems or software versions, and
adaptation of the approach to other rate-based constitutive models,
the full derivation of the analytical solution is provided in the
\nameref{sec:Appendix}, and simulation input files and post-processing
scripts are provided as supplemental material.

\section*{Conflict of interest statement}

One of the authors (NR) was formerly an employee of Dassault Systèmes
Simulia, makers of \texttt{ABAQUS}.

\section*{Acknowledgments}

\noindent We thank Snehal S. Shetye and Andrew P. Baumann (U.S. FDA)
for reviewing the manuscript. This study was funded by the U.S. FDA
Center for Devices and Radiological Health (CDRH) Critical Path program.
The research was supported in part by an appointment to the Research
Participation Program at the U.S. FDA administered by the Oak Ridge
Institute for Science and Education through an interagency agreement
between the U.S. Department of Energy and FDA. The findings and conclusions
in this article have not been formally disseminated by the U.S. FDA
and should not be construed to represent any agency determination
or policy. The mention of commercial products, their sources, or their
use in connection with material reported herein is not to be construed
as either an actual or implied endorsement of such products by the
Department of Health and Human Services.

\clearpage{}

\appendix

\section{Appendix\label{sec:Appendix}}

In the following, standard typeface symbols are scalars (e.g., $\phi$),
boldface symbols denote second-order tenors (e.g., $\mathbf{F}$ or
$\boldsymbol{\sigma}$), and blackboard bold characters denote fourth-order
tensors (e.g., $\mathbb{A}$). Additionally, the overdot operator
(e.g., $\dot{A}$) indicates a time derivative.

\subsection{General kinematic equations}

Begin with the problem setup provided in Section \ref{subsec:Problem-setup}
of the manuscript. The deformation gradient tensor is
\begin{equation}
\mathbf{F}=\mathbf{I}+\nabla\mathbf{u}
\end{equation}
or, in matrix form,
\begin{equation}
\mathbf{F}=\begin{bmatrix}\lambda & 0 & 0\\
0 & 1 & 0\\
0 & 0 & 1
\end{bmatrix}
\end{equation}
where $\mathbf{I}$ is the second-order identity tensor and $\lambda=\frac{l}{L_{0}}$
is the stretch. For finite-strain problems, \texttt{ABAQUS} solves
the problem incrementally and calculates the strain by integrating
the strain increments. The resulting strain measure is thus the logarithmic
or ``true'' strain 
\begin{equation}
\boldsymbol{\epsilon}=\ln\mathbf{V}
\end{equation}
where $\ln$ is the principal matrix logarithm and 
\begin{equation}
\mathbf{V}=\sqrt{{\mathbf{F}\mathbf{F}}^{T}}
\end{equation}
is the left Cauchy stretch tensor.

Since $\mathbf{F}$ is diagonal here, $\mathbf{F}=\mathbf{F}^{T}$
and $\mathbf{V}=\mathbf{F}$. Therefore, the logarithmic strain tensor
is simply $\ln\mathbf{F}$ or 
\begin{equation}
\boldsymbol{\epsilon}=\begin{bmatrix}\ln\lambda & 0 & 0\\
0 & 0 & 0\\
0 & 0 & 0
\end{bmatrix}\,.
\end{equation}
Since the only non-zero strain component for this uniaxial strain
problem is $\epsilon_{11}$, let 
\begin{align}
\epsilon & =\epsilon_{11}\nonumber \\
 & =\ln\lambda\label{eqn:eps}
\end{align}
to simplify notation.

The volumetric strain $\epsilon_{V}$ is next defined as 
\begin{align}
\epsilon_{V} & =\mathrm{tr}\,\boldsymbol{\epsilon}\nonumber \\
 & =\epsilon_{11}+\epsilon_{22}+\epsilon_{33}\nonumber \\
 & =\epsilon\,,\label{eq:volumetric-strain}
\end{align}
which represents a measure of volume change or dilation, where $\textrm{tr}\,\mathbf{A}=\mathbf{A}:\mathbf{I}$
is the trace operator for a second-order tensor and $\left(:\right)$
is the double-inner product operator $\mathbf{A}:\mathbf{B}=A_{ij}B_{ij}$.
Note that the volumetric strain is sometimes defined in other literature
as $\epsilon_{V}=\frac{1}{3}\textrm{tr}\,\boldsymbol{\epsilon}$,
which represents a measure of mean normal strain.

The strain rate tensor 
\begin{equation}
\dot{\boldsymbol{\epsilon}}=\begin{bmatrix}\dot{\epsilon} & 0 & 0\\
0 & 0 & 0\\
0 & 0 & 0
\end{bmatrix}
\end{equation}
is obtained by taking the time derivative of $\boldsymbol{\epsilon}$.
The deviatoric strain rate tensor $\dot{\mathbf{e}}$ is 
\begin{equation}
\dot{\mathbf{e}}=\boldsymbol{\dot{\epsilon}}-\frac{1}{3}\,\textrm{tr}\,\dot{\boldsymbol{\epsilon}}\,\mathbf{I}\,,\label{eq:dev-strain}
\end{equation}
or, in matrix form,
\begin{equation}
\dot{\mathbf{e}}=\begin{bmatrix}\frac{2}{3}\dot{\epsilon} & 0 & 0\\
0 & -\frac{1}{3}\dot{\epsilon} & 0\\
0 & 0 & -\frac{1}{3}\dot{\epsilon}
\end{bmatrix}\,.\label{eq:deviatoric-strain-rate-matrix}
\end{equation}
The scalar equivalent strain rate $\dot{e}$ is 
\begin{align}
\dot{e} & =\sqrt{\frac{2}{3}\dot{\mathbf{e}}:\dot{\mathbf{e}}}\nonumber \\
 & =\frac{2}{3}\dot{\epsilon}\,.\label{eqn:plst-eq-strain}
\end{align}

\subsection{Equations for linear, monotonic transformation behavior}

The superelastic constitutive model in \texttt{ABAQUS} \cite{rebelo2000finite,rebelo2001simulation}
is based on the work of Aurrichio and Taylor \cite{auricchio1997shape,auricchioTaylorLubliner1997}
and leverages generalized plasticity theory to model the dependency
of the material stiffness on the current stress state (see Online
SIMULIA User Assistance 2022 \textgreater Abaqus \textgreater Materials
\textgreater Elastic Mechanical Properties \textgreater Superelasticity).
The constitutive model uses the additive strain rate decomposition
\begin{equation}
\dot{\boldsymbol{\epsilon}}^{e}=\dot{\boldsymbol{\epsilon}}-\dot{\boldsymbol{\epsilon}}^{\textrm{tr}}\,,\label{eq:strain-decomp}
\end{equation}
where $\dot{\boldsymbol{\epsilon}}^{e}$ is the elastic strain rate
tensor and $\dot{\boldsymbol{\epsilon}}^{\textrm{tr}}$ is the transformation
strain rate tensor. The Cauchy stress rate tensor is then 
\begin{equation}
\dot{\boldsymbol{\sigma}}=\mathbb{D}:\dot{\boldsymbol{\epsilon}}^{e}\label{eq:Cauchy-stress-rate}
\end{equation}
where $\mathbb{D}$ is the fourth-order elasticity or stiffness tensor
and $\left(:\right)$ is the operator $\mathbb{A}:\mathbf{B}=A_{ijkl}B_{kl}$.
For an isotropic material,

\begin{equation}
\mathbb{D}=2G\,\left(\mathbb{I}-\frac{1}{3}\,\mathbf{I}\otimes\mathbf{I}\right)+K\,\mathbf{I}\otimes\mathbf{I}\,,\label{eq:Fourth-order-stiffness}
\end{equation}
where $G$ is the shear modulus, $K$ is the bulk modulus, $\left(\mathbf{I}\otimes\mathbf{I}\right)_{ijkl}=\delta_{ij}\delta_{kl}$,
$\left(\mathbb{I}\right)_{ijkl}=\frac{1}{2}\left(\delta_{ik}\delta_{jl}+\delta_{il}\delta_{jk}\right)$,
and $\delta_{ij}$ is the Kronecker delta function $\delta_{ij}=\begin{cases}
0 & i\neq j\\
1 & i=j
\end{cases}$ \cite{de2011computational}. Substituting Eqn. \ref{eq:Fourth-order-stiffness}
into Eqn. \ref{eq:Cauchy-stress-rate} and simplifying, the Cauchy
stress rate can be written in Hooke's law form as 
\begin{align}
\dot{\boldsymbol{\sigma}} & =2G\left(\dot{\boldsymbol{\epsilon}}^{e}-\frac{1}{3}\textrm{tr}\,\dot{\boldsymbol{\epsilon}}^{e}\mathbf{I}\right)+K\,\textrm{tr}\,\dot{\boldsymbol{\epsilon}}^{e}\mathbf{I}\nonumber \\
 & =2G\dot{\mathbf{e}}^{e}+K\,\textrm{tr}\,\dot{\boldsymbol{\epsilon}}^{e}\mathbf{I}\,.\label{eq:Cauchy-stress-Hookean}
\end{align}

In \texttt{ABAQUS}, the flow rule describing the transformation strain
rate for superelastic materials is 
\begin{equation}
\dot{\boldsymbol{\epsilon}}^{\textrm{tr}}=\epsilon^{L}\,\dot{\xi}\,\frac{\partial G^{\textrm{tr}}}{\partial\boldsymbol{\sigma}}\,,\label{eqn:eps-tr-1}
\end{equation}
where $\epsilon^{L}$ is a material constant, $\xi$ is the martensite
fraction, and $G^{\textrm{tr}}$ is a Drucker--Prager type transformation
potential 
\begin{equation}
G^{\textrm{tr}}=q-p\tan\psi\,.
\end{equation}
In the above, $p$ is hydrostatic pressure, $\psi$ is a scaling constant,
and $q$ is the von Mises stress
\begin{equation}
q=\sqrt{\frac{3}{2}\mathbf{S}:\mathbf{S}}\,,\label{eq:von-Mises}
\end{equation}
where $\mathbf{S}$ is the deviatoric stress tensor

\begin{equation}
\mathbf{S}=\boldsymbol{\sigma}-\frac{1}{3}\,\textrm{tr}\boldsymbol{\sigma}.\label{eq:deviatoric-stress-tensor}
\end{equation}
As stated earlier, here we assume symmetric compression and tension
behavior (i.e., $\sigma_{L}^{S}=\sigma_{cL}^{S}$). Thus, $\psi=0$,
and the transformation potential takes on a von Mises form as simply
\begin{equation}
G^{\textrm{tr}}=q\,.\label{eq:Gtr-reduced}
\end{equation}

\subsection{An aside: radial return and the direction tensor $\mathbf{n}$}

Because the transformation behavior has been simplified as von Mises-type,
the deviatoric stress and deviatoric strain rate tensors point in
the same direction (in 6D space). The transformation potential is
$q$, and the transformation strain rate is proportional to (i.e.,
in the direction of) the gradient of $q$ with respect to stress $\frac{\partial q}{\partial\boldsymbol{\sigma}}$,
which we denote as $\mathbf{n}$. Expanding using Eqn. \ref{eq:von-Mises}
and applying chain rule,
\begin{align}
\mathbf{n} & =\frac{\partial q}{\partial\boldsymbol{\sigma}}\nonumber \\
 & =\frac{\partial\sqrt{\frac{3}{2}\mathbf{S}:\mathbf{S}}}{\partial\boldsymbol{\sigma}}\nonumber \\
 & =\frac{\partial\left(\mathbf{S}:\mathbf{S}\right)^{\frac{1}{2}}}{\partial\mathbf{S}}\cdot\frac{\partial\mathbf{S}}{\partial\boldsymbol{\sigma}}\nonumber \\
 & =\frac{1}{2}\left(\frac{3}{2}\mathbf{S}:\mathbf{S}\right)^{-\frac{1}{2}}\frac{3}{2}\left(2\mathbf{S}\right)\cdot\frac{\partial(\boldsymbol{\sigma}-\frac{1}{3}\,\textrm{tr}\boldsymbol{\sigma})}{\partial\boldsymbol{\sigma}}\nonumber \\
 & =\frac{3}{2}\frac{\mathbf{S}}{q}\cdot(\mathbb{I}-\frac{1}{3}\,\mathbf{I}\otimes\mathbf{I})\nonumber \\
 & =\frac{3}{2}\frac{\mathbf{S}}{q}\,.\label{eq:S-and-n}
\end{align}
Note the inner product of $\mathbf{n}$ with itself is
\begin{equation}
\begin{aligned}[b]\mathbf{n}:\mathbf{n} & =\frac{3}{2}\left(\frac{3}{2}\frac{\mathbf{S}:\mathbf{S}}{q^{2}}\right)\\
 & =\frac{3}{2}\frac{q^{2}}{q^{2}}\\
 & =\frac{3}{2}\,.
\end{aligned}
\label{eq:dbl-prd-n}
\end{equation}
The double-inner product between the deviatoric stress tensor $\mathbf{S}$
and the direction of the deviatoric strain $\mathbf{n}$ is also useful
since
\begin{align}
\mathbf{S}:\mathbf{n} & =\frac{3}{2}\frac{\mathbf{S}:\mathbf{S}}{q}\nonumber \\
 & =\frac{q^{2}}{q}\nonumber \\
 & =q\,.\label{eq:q-von-Mises}
\end{align}

Radial return algorithms project the trial stress back onto the yield
(here transformation) surface by scaling the stress radially with
respect to the hydrostatic axis $\sigma_{1}=\sigma_{2}=\sigma_{3}$,
where $\sigma_{i}$ are principal stresses. We assume radial return
to be exact under the given simplifications and approximations, specifically,
equal transformation stresses in compression and tension and thereby
von Mises-like transformation behavior, and proportional (radial)
loading. Accordingly, the loading direction coincides with the projection
direction and the (pseudo)plastic strain rate direction. As derived
below,
\begin{equation}
\dot{\mathbf{e}}^{tr}=\dot{e}^{tr}\,\mathbf{n}\label{eq:dot-e}
\end{equation}
where $\dot{\mathbf{e}}^{tr}$and $\dot{e}^{tr}$ are the deviatoric
transformation strain rate tensor and equivalent scalar, respectively,
and $\mathbf{n}$ specifies the direction of the deviatoric transformation
strain rate (the normal direction to the transformation surface with
the given problem description). Following standard Mises plasticity
arguments and given proportional loading, $\dot{\mathbf{e}}$ and
$\dot{\mathbf{e}}^{tr}$are collinear. Accordingly, using Eqn. \ref{eqn:plst-eq-strain},
the deviatoric strain rate tensor may also be written
\begin{align}
\dot{\mathbf{e}} & =\dot{e}\,\mathbf{n}\nonumber \\
 & =\frac{2}{3}\dot{\epsilon}\,\mathbf{n}\label{eq:dot-e-1}
\end{align}
and therefore, given Eqn. \ref{eq:deviatoric-strain-rate-matrix},
\begin{equation}
\mathbf{n}=\begin{bmatrix}1 & 0 & 0\\
0 & -\frac{1}{2} & 0\\
0 & 0 & -\frac{1}{2}
\end{bmatrix}\,.
\end{equation}

\subsection{Equations for linear, monotonic transformation behavior (continued)}

Continuing from Eqn. \ref{eqn:eps-tr-1}, using Eqn. \ref{eq:Gtr-reduced}
and the definition of $\mathbf{n}$, the last term on the right-hand
side becomes
\begin{align}
\frac{\partial G^{\textrm{tr}}}{\partial\boldsymbol{\sigma}} & =\frac{\partial q}{\partial\boldsymbol{\sigma}}\nonumber \\
 & =\mathbf{n}\,.\label{eqn:Gtr-reduced-vector-notation}
\end{align}
The transformation strain rate from Eqn. \ref{eqn:eps-tr-1} can thus
be written
\begin{equation}
\dot{\boldsymbol{\epsilon}}^{\textrm{tr}}=\epsilon^{L}\,\dot{\xi}\,\mathbf{n}\,,\label{eq:transformation-strain}
\end{equation}
and the scalar equivalent transformation strain rate becomes
\begin{align}
\dot{\epsilon}^{\textrm{tr}} & =\sqrt{\frac{2}{3}\dot{\boldsymbol{\epsilon}}^{\textrm{tr}}:\dot{\boldsymbol{\epsilon}}^{\textrm{tr}}}\nonumber \\
 & =\epsilon^{L}\,\dot{\xi}\,.\label{eq:transformation-strain-1}
\end{align}

Note that $\mathbf{n}$ is deviatoric and $\textrm{tr}\,\mathbf{n}=0$.
Therefore, 
\begin{equation}
\textrm{tr}\,\dot{\boldsymbol{\epsilon}}^{\textrm{tr}}=0
\end{equation}
and 
\begin{align}
\textrm{tr}\,\dot{\boldsymbol{\epsilon}}^{e} & =\textrm{tr}\,\left(\dot{\boldsymbol{\epsilon}}-\dot{\boldsymbol{\epsilon}}^{\textrm{tr}}\right)\nonumber \\
 & =\textrm{tr}\,\dot{\boldsymbol{\epsilon}}\nonumber \\
 & =\dot{\epsilon}_{V}\,,\label{eq:tr-eps-e-dot}
\end{align}
where $\dot{\epsilon}_{V}$ is the volumetric strain rate.

Using Eqns. \ref{eq:dev-strain}, \ref{eq:strain-decomp}, and \ref{eq:transformation-strain}---\ref{eq:tr-eps-e-dot},
the Cauchy stress rate from Eqn. \ref{eq:Cauchy-stress-Hookean} becomes
\begin{align}
\dot{\boldsymbol{\sigma}} & =2G\left[\left(\dot{\boldsymbol{\epsilon}}-\epsilon^{L}\,\dot{\xi}\,\mathbf{n}\right)-\frac{1}{3}\textrm{tr}\,\left(\dot{\boldsymbol{\epsilon}}-\epsilon^{L}\,\dot{\xi}\,\mathbf{n}\right)\mathbf{I}\right]+K\,\dot{\epsilon}_{V}\mathbf{I}\nonumber \\
 & =2G\left[\left(\dot{\boldsymbol{\epsilon}}-\epsilon^{L}\,\dot{\xi}\,\mathbf{n}\right)-\frac{1}{3}\textrm{tr}\,\left(\dot{\boldsymbol{\epsilon}}\right)\mathbf{I}\right]+K\,\dot{\epsilon}_{V}\mathbf{I}\nonumber \\
 & =2G\left(\dot{\mathbf{e}}-\epsilon^{L}\,\dot{\xi}\,\mathbf{n}\right)+K\,\dot{\epsilon}_{V}\,\mathbf{I}\,.
\end{align}
The Cauchy stress rate can be split into deviatoric and hydrostatic
components
\begin{equation}
\dot{\boldsymbol{\sigma}}=\dot{\mathbf{S}}-\dot{p}\,\mathbf{I}\,,\label{eq:Cauchy-split}
\end{equation}
where
\begin{equation}
\dot{\mathbf{S}}=2G\left(\dot{\mathbf{e}}-\epsilon^{L}\,\dot{\xi}\,\mathbf{n}\right)\label{eq:deviatoric-stress-2}
\end{equation}
is the deviatoric stress rate and
\begin{equation}
\dot{p}=-K\,\dot{\epsilon}_{V}\label{eq:hydrostatic-stress-2}
\end{equation}
is the hydrostatic pressure rate.

The martensite fraction rate $\dot{\xi}$ is a function of the equivalent
stress,
\begin{equation}
\dot{\xi}=f(q)\,,
\end{equation}
and this function is called the transformation law (equivalent to
a work-hardening law in plasticity). In general, the martensite fraction
rate must be calculated for each increment, as its sign and magnitude
depend on the current stress state as well as the direction and magnitude
of the stress rate. However, for the linear hardening approximation
and monotonic loading, we can define the martensite fraction directly
as a linear function of the von Mises equivalent stress. For loading,
we have 
\begin{equation}
\xi_{\textrm{L}}\left(q\right)=\begin{cases}
0 & q\leq\sigma_{L}^{S}\\
\frac{q-\sigma_{L}^{S}}{\sigma_{L}^{E}-\sigma_{L}^{S}} & \sigma_{L}^{S}\lt q\lt\sigma_{L}^{E}\\
1 & q\geq\sigma_{L}^{E}
\end{cases}\,,
\end{equation}
and for unloading, 
\begin{equation}
\xi_{\textrm{U}}\left(q\right)=\begin{cases}
0 & q\leq\sigma_{U}^{E}\\
\frac{q-\sigma_{U}^{E}}{\sigma_{U}^{S}-\sigma_{U}^{E}} & \sigma_{U}^{E}\lt q\lt\sigma_{U}^{S}\\
1 & q\geq\sigma_{U}^{S}
\end{cases}\,.
\end{equation}

For strictly monotonic proportional (constant $\mathbf{n}$) loading
or unloading, we can now integrate the rate tensors in Eqns. \ref{eq:Cauchy-split}---\ref{eq:hydrostatic-stress-2}
from $t=0$ to $t=t$ to obtain 
\begin{equation}
\boldsymbol{\sigma}=\mathbf{S}-p\,\mathbf{I}\label{eqn:sigma}
\end{equation}
\begin{equation}
\mathbf{S}=2G\,\left(\mathbf{e}-\epsilon^{L}\,\xi\,\mathbf{n}\right)\label{eqn:S-dev}
\end{equation}
\begin{equation}
p=-K\,\epsilon_{V}\,.\label{eqn:p}
\end{equation}
Similarly, substituting Eqn. \ref{eqn:S-dev} into Eqn. \ref{eq:q-von-Mises}
and expanding using Eqns. \ref{eq:dot-e-1} and \ref{eq:dbl-prd-n},
the equivalent stress $q$ becomes 
\begin{align}
q & =\mathbf{S}:\mathbf{n}\nonumber \\
 & =2G\left(\mathbf{e}:\mathbf{n}-\epsilon^{L}\,\xi\,\mathbf{n}:\mathbf{n}\right)\nonumber \\
 & =2G\left(\frac{2}{3}\epsilon\,\mathbf{n}:\mathbf{n}-\epsilon^{L}\,\xi\,\mathbf{n}:\mathbf{n}\right)\nonumber \\
 & =2G\left(\epsilon-\frac{3}{2}\epsilon^{L}\,\xi\right)\nonumber \\
 & =2G\,\epsilon-3G\epsilon^{L}\,\xi\,.\label{eqn:q}
\end{align}
Note the integrated equations take the same form for both loading
and unloading. Pairings of $q$ and $\xi$, however, define specific
locations on the stress-strain curve such that the solutions for loading
and unloading are unique (see Figure \ref{fig:NiTi-linear-vs-cubic-transformation}
and Table \ref{tab:verification-points}).

\subsection{Final analytical solution for uniaxial strain conditions}

Given $q$ and $\xi$ (Table \ref{tab:verification-points}), solve
Eqn. \ref{eqn:q} for the total logarithmic strain $\epsilon$,
\begin{equation}
\boxed{\epsilon=\underbrace{\frac{q}{2G}}_{\text{elastic strain}}+\underbrace{\frac{3}{2}\epsilon^{L}\xi}_{\text{transformation strain}}}\,.
\end{equation}
Next, use Eqn. \ref{eq:S-and-n} to calculate the deviatoric stress
tensor,
\begin{equation}
\boxed{\mathbf{S}=\frac{2}{3}q\,\mathbf{n}}
\end{equation}
and Eqns. \ref{eqn:p} and \ref{eq:volumetric-strain} to calculate
the hydrostatic pressure,

\begin{equation}
\boxed{p=-K\epsilon}\,.
\end{equation}
Use Eqn. \ref{eqn:sigma} to calculate the Cauchy stress tensor,

\begin{equation}
\boxed{\boldsymbol{\sigma}=\mathbf{S}-p\mathbf{I}}\,.
\end{equation}
Finally, using Eqns. \ref{eqn:stretch} and \ref{eqn:eps}, calculate
the corresponding solution pseudo-time $t$,

\begin{equation}
\boxed{t=\begin{cases}
\frac{L_{0}\left(\lambda-1\right)}{u} & \text{loading}\\
1-\frac{L_{0}\left(\lambda-1\right)}{u} & \text{unloading}
\end{cases}}
\end{equation}
where $\lambda=\exp\left(\epsilon\right)$.

\noindent \clearpage{}

\bibliographystyle{IEEEtran}
\bibliography{MES_references}

\end{document}